\documentclass[letterpaper, 10 pt, conference]{ieeeconf}  % Comment this line out if you need a4paper

\IEEEoverridecommandlockouts                              % This command is only needed if 
\usepackage{graphicx}
\usepackage{subfigure}
\usepackage{comment}
\usepackage{multirow}
\usepackage{xcolor}
\usepackage{tcolorbox}
\usepackage{dirtytalk}
\usepackage{quoting}
\usepackage[hidelinks]{hyperref}

\title{\LARGE \bf
A Robot That Listens: Enhancing Self-Disclosure and Engagement Through Sentiment-based Backchannels and Active Listening
}

\author{
    Hieu Tran$^{*1}$, Go-Eum Cha$^{1}$, Sooyeon Jeong$^{1}$
    \thanks{$^{*}$ Corresponding author: {\tt\small tran335@purdue.edu}}
    \thanks{$^{1}$ Department of Computer Science, Purdue University, West Lafayette, USA}
}

\begin{document}

\maketitle
% \begin{figure*}[t]
%     \centering
%     \subfigure[\textit{Control}]{\includegraphics[scale=0.109]{figures/demo_1.jpg}}
%     \subfigure[\textit{Backchanneling-only} ]{\includegraphics[scale=0.109]{figures/demo_2.jpg}} 
%     \subfigure[\textit{Backchanneling+Active Listening} ]{\includegraphics[scale=0.109]{figures/demo_3.jpg}}
%     \caption{We investigated the impact of three robot listening behavior on people's perceived empathy, closeness, and self-disclosures. The three robot listening behaviors we explored were: (a) the control behavior (\textit{control}), in which the robot only asked questions, (b) the backchanneling-only behavior (\textit{BC}), in which the robot produced sentiment-based backchannels while listening to the participant's response, and (c) the backchanneling+active listening behavior (\textit{BC+AL}), in which the robot not only backchanneled but also responded with an active listening manner.}
%     \label{fig:three_condition_demo}
% \end{figure*}

\thispagestyle{empty}
\pagestyle{empty}

%%%%%%%%%%%%%%%%%%%%%%%%%%%%%%%%%%%%%%%%%%%%%%%%%%%%%%%%%%%%%%%%%%%%%%%%%%%%%%%%

\begin{abstract}

As social robots get more deeply integrated into our everyday lives, they will be expected to engage in meaningful conversations and exhibit socio-emotionally intelligent listening behaviors when interacting with people. %In this paper, we present a social robot that can generate sentiment-based backchannels and active listening behaviors during its conversations with human interlocutors. 
Active listening and backchanneling could be one way to enhance robots' communicative capabilities and enhance their effectiveness in eliciting deeper self-disclosure, providing a sense of empathy, and forming positive rapport and relationships with people. Thus, we developed an LLM-powered social robot that can exhibit contextually appropriate sentiment-based backchanneling and active listening behaviors (\textit{active listening+backchanneling}) and compared its efficacy in eliciting people's self-disclosure in comparison to robots that do not exhibit any of these listening behaviors (\textit{control}) and a robot that only exhibits backchanneling behavior (\textit{backchanneling-only}). Through our experimental study with sixty-five participants, %; (1) \textit{control} behavior that only asked questions and passively listened to participant's responses, (2) a \textit{backchanneling-only} behavior, in which the robot produced sentiment-based backchannels, and (3) an \textit{active listening + backchanneling behavior}, in which the robot not only sentiment-based backchanneled but also exhibited active listening behavior in terms of longer utterances. Sixty-five participants were randomized into one of the three experimental conditions and conversed with a Furhat robot based on a set of questions designed to elicit varying levels of self-disclosures. 
we found the participants who conversed with the active listening robot perceived the interactions more positively, in which they exhibited the highest self-disclosures, 
% especially for sensitive topics, 
and reported the strongest sense of being listened to.  %We also notice that %In addition, participants perceived some level of empathy in the active listening condition. 
The results of our study suggest that the implementation of active listening behaviors in social robots has the potential to improve human-robot communication and could further contribute to the building of deeper human-robot relationships and rapport. 

\end{abstract}

\section{Introduction}

Mutual and reciprocal self-disclosure is one way for social robots to build trust and relationships with people \cite{wheeless1977measurement,masaviru2016self, hill2001self}. %For instance, Jeong et al. \cite{jeong2023robotic} found that people who interacted with a robot that shows self-disclosure behavior showed better psychological well-being outcomes and reported stronger working alliance with the robot than people who interacted with a robot that only provides instructions for the tasks. 
Eliciting people's self-disclosure is crucial when the role of interactive robots is to serve intimate support roles in various application contexts (e.g., healthcare, elder care, or education). With the advent of Large Language Models (LLMs), social robots have opportunities to provide highly personalized and tailored interactions that result in high engagement~\cite{kim2024,lozano2024}, trust~\cite{anzabi2023effect}, and perceived empathy~\cite{lee2023developing} between humans and robots~\cite{pasternak2023}. 
%Yet, LLM-powered social robots still have limitations in meeting people's expectations in sophisticated embodied conversation and fall short in enacting appropriate non-verbal cues and building connection \cite{kim2024understanding}.
 %It is also shown benefit physical and mental health  \cite{rosenberg2002expressive,smyth2012health}. 

One way to elicit self-disclosure in dyadic interactions is active listening~\cite{levitt2002active}, which refers to the process of being fully attentive to the speaker, noticing what and how they say it, and responding in a way that shows you understand and are engaged with their messages~\cite{jones2013communication}. Engaging in active listening is shown to elicit deep self-disclosure and intimacy, resulting in improved interpersonal relationships and trust \cite{weger2014relative,hutchby2005active,jones2011supportive,kuhn2018power}. Several prior works investigated the effect of conversational agents' active listening and/or backchanneling on people's perception of the robots. However, most existing works on agent backchanneling only focus on neutral backchannels (e.g., ``mmhmm'' and ``I see'') but do not include sentimented backchannels (e.g., ``oh wow!'', ``oh no...'', ``goodness!'') and little work has explored how a robot's listening behavior influences people's self-disclosures and rapport with the robot. %In addition, many of these prior works were conducted prior to recent development of LLMs, which can further enhance the robot's capability to tailor and personalize its utterances in its active listening behavior. %It can also help promote empathetic conversation between human and human. 

Thus, we present an autonomous LLM-powered robot that can exhibit \textbf{sentiment-based backchanneling} and \textbf{active listening} behavior to investigate whether these attentive and empathetic listening behaviors elicit deep self-disclosures and engagement from people during a conversational task.
%In particular, we focused on the listening module of the robots. Active listening is the key to foster the relationship between humans in the first encounter~\cite{weger2014relative}. 
%Throughout this conversation, based on the speakers' messages, o
%We developed three different robot listening behaviors that (1) listen-only (\textit{control} condition), (2) backchannels with affect/emotion (\textit{BC} condition), and (3) actively listens and backchannels (\textit{BC+AL} condition), and conducted an experimental study to compare the effects of these behaviors on people's perceived empathy, closeness, and the quantity and quality of their self-disclosures. 
Through this investigation, we aim to deepen the understanding of how a robot's listening behavior influences people's willingness to disclose personal information and emotions about themselves and how it could impact the overall interaction experience and rapport with the robot. With the results from this study, we hope to offer valuable insights into the development of social robots that can provide more personalized support and a positive human-robot relationship.

\section{Related Work} 
\subsection{Active Listening and Backchanneling}
The experience of being heard and accepted is a crucial component of social communication and can be achieved when a listener exhibits ``good listening behavior~\cite{lavee2023good, rogers1995way}.'' Good listening behavior is the ability to sincerely and actively comprehend the speaker's perspectives, which is characterized by empathy, non-judgment, and an attitude of respect and curiosity toward the speaker, is shown to enhance trust, rapport, and relationship between the speaker and the listener~\cite{castro2016does, itani2019building, itzchakov2020can}. 

Backchanneling and Active Listening are two components that comprise good listening behavior. Backchanneling is widely used in various languages in everyday life (e.g., English, Japanese, or Chinese) with slight variation depending on the contexts and languages~\cite{heinz2003backchannel, li2010backchannel, senk1997analyzing}. %It includes non-lexical utterances and various gestures (e.g., nodding or shaking head) and ~\cite{park2017telling}. 
People often use generic backchannels (e.g., \textit{yeah}, \textit{ok}, \textit{uh-huh}, or \textit{mm-hmm}) to signal to the speaker that they are listening, but also use specific backchannel tailored to the context of the ongoing conversation (e.g., \textit{Oh no ...} when someone shares a struggle). For instance, a study by Bavelas et al.~\cite{bavelas2000listeners} found that people performed better in a storytelling task when a listener exhibited specific and contextually appropriate backchannels compared to when they received generic responses. 

Active listening is another psychotherapeutic approach to convey unconditional acceptance and acknowledgment of the speaker's experience~\cite{orlov1992, rogers2012client}. In active listening, the listener rephrases or repeats the speaker's messages to validate what they are conveying~\cite{weger2010active, garland1981training} and is frequently followed by asking questions that encourage the speaker to further engage in the conversation~\cite{paukert2004}. These behaviors make the speakers feel unconditional acceptance and build trust from the listener, especially during the first encounter~\cite{rogers2012client, weger2010active}. 

\raggedbottom
\subsection{Developing Attentive Behaviors for Interactive Agents}
Several prior studies investigated ways to implement attentive or good listening behaviors in robots or artificial agents and how these behaviors impact human-robot/agent interactions and people's perception of the interactive agents. % Since active listening can positively influence communicative satisfaction from enhanced understanding, leading to endorsing a sense of being validated \cite{weger2014relative}, several studies aimed to implement attentive listening behaviors on robots and artificial agents. 
Among them, some focused on predicting the appropriate timing for a robot's backchannel behavior. Murray et al.~\cite{murray2022learning} developed a long short-term memory (LSTM) neural network model that allowed virtual agents to perform more human-like head-nodding. Hussain et al.~\cite{hussain2022} and Gillet et al.~\cite{gillet2024shielding} used reinforcement learning methods to identify appropriate timing for nodding and smiles. Others used deep learning methods to predict the appropriate timing of the nodding and smiles~\cite{murray2022learning, hussain2022, gillet2024shielding}. 

Others investigated the effect of well-timed backchannel behaviors on the interaction experience and perception of the robots. Park et al.~\cite{park2017backchannel, park2017telling} developed a rule-based backchanneling opportunity prediction (BOP) system and found that children preferred interacting with a robot that exhibited attentive backchannel behavior over a robot with non-contingent backchannel behavior. Similarly, Cho et al.~\cite{cho2022alexa} demonstrated that pseudo-random backchannels incorporated in Alexa devices improved people's perceived emotional support and elicited more self-disclosures. 
%On the other hand, several works have utilized the ruled-based and random method to mimic human backchannel behaviors on agents~\cite{park2017backchannel, park2017telling, cho2022alexa}. 

\raggedbottom

However, most prior works focused on generating or predicting timing for generic/neutral backchannels (e.g., ``mm-hmm'', ``uh-huh'') or gestures (e.g., nodding), and do not explore the effect of contextually specific or sentiment-based robot backchannels. In Shaverdi et al.'s work~\cite{shahverdi2023}, humans were displaying a variety of specific emotional responses (i.e., positive, neutral, and negative) driven by the stories, but this work was conducted in a Wizard-of-Oz study and was not an autonomous system. %of the Wizard-of-OZ (WOZ) social robots. Thus, this work highlighted the importance of emotional factors in human and social robot interactions.
Similarly, Anzabi et al.~\cite{anzabi2023effect} conducted a WoZ study to investigated the effect of non-verbal and verbal active listening traits (e.g., generic backchanneling, summarizing, and asking for questions) on perceived empathy and trust. 

With the recent development of LLM, robots and chat interfaces can perform more human-like conversations with empathetic responses via zero-shot learning or few-shot learning~\cite{pasternak2023, chen2023soulchat, qian2023harnessing, lee2022does}. For instance, Pasternak et al.~\cite{pasternak2023} demonstrated that LLMs-prompted active listening robots enhanced users' perception of the robot's social behavior rather than scripted communication. Several works demonstrated the empathetic responses via chat interface~\cite{chen2023soulchat, qian2023harnessing, lee2022does}.

Building upon these prior works, we developed an autonomous robot system that generated adaptive active listening behaviors and sentiment-based backchannels. Our system utilized both linguistic and acoustic features of people's verbal utterances. This approach helped achieve more humanistic responses by more flexible systems and enhanced the intimacy between humans and social robots. To the best of our knowledge, no previous work studied the effect of context-based backchanneling and responses as a whole.

\begin{figure*}[ht!]
    \centering
    \includegraphics[width=\linewidth]{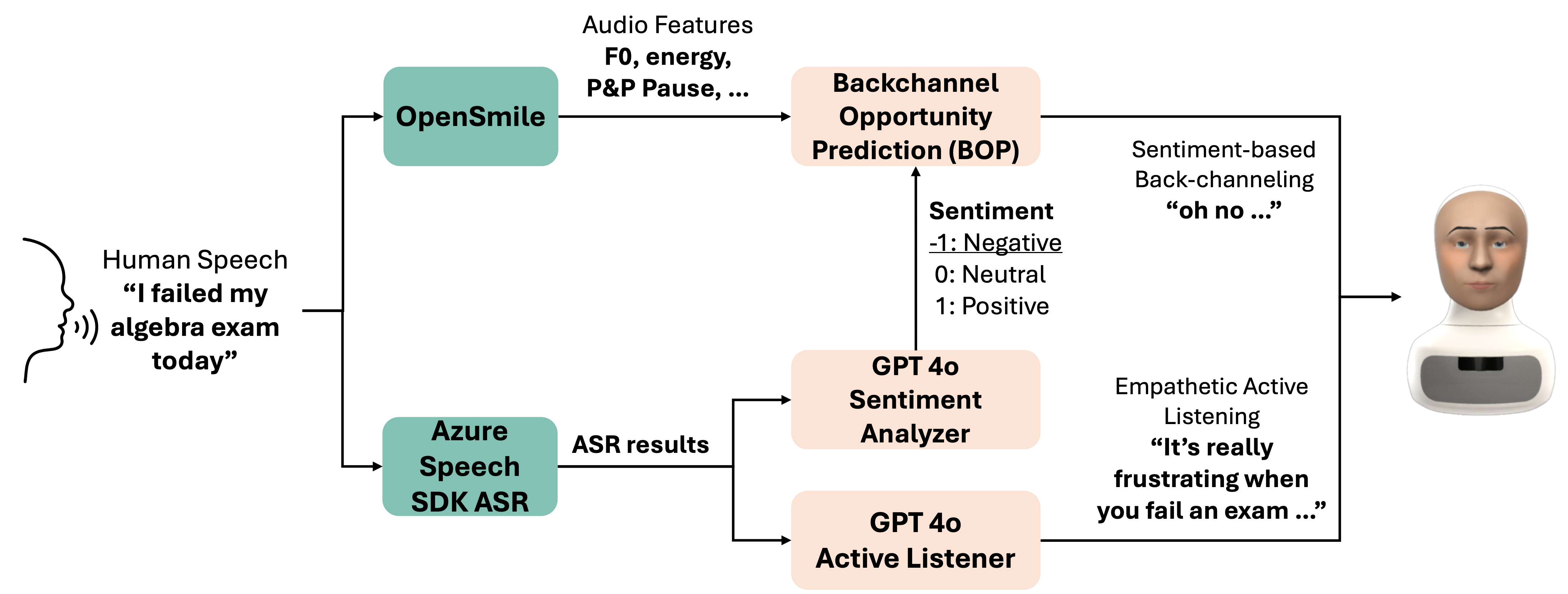}
    \caption{An overview of system architecture. When a participant answers a question, the prosodic features were extracted via {OpenSmile} \cite{park2017backchannel} and the verbal utterances were recognized via the \textit{Azure Speech SDK ASR} module. The \textit{BOP} module identified the appropriate timing for the robot to backchannel and the \textit{Sentiment Analyzer} module classified the participant's utterance as positive, neutral, or negative to generate sentiment-based backchanneling behavior for the robot. The \textit{Active Listener} module used the \textit{Azure Speech SDK ASR} results to generate empathetic robot responses based on the prompted active listening principles.}
    \label{fig:sys_architecture}
\end{figure*}

\section{Developing a Robotic Active Listener}
This section outlines how we designed and implemented our LLM-powered listening behavior for the robot. Our system consisted of two core components: (1) \textit{Sentiment-based Backchanneling} and (2) \textit{Active Listening}. The \textit{Sentiment-based Backchanneling} module used both acoustic/prosodic features to determine the appropriate timing for the robot to backchannel and the sentiment of the participant's utterances to determine the type of sentiment-based backchannels to generate (i.e. positive, neutral, or negative). The \textit{Active Listening} module generated empathetic responses, which is tailored to each participant's answers, from the prompt-engineered LLM (GPT-4o) output (Figure~\ref{fig:sys_architecture}). Our system was implemented on the Furhat robot to converse with study participants. 

% \footnote{\url{https://learn.microsoft.com/en-us/azure/ai-services/speech-service/speech-sdk}}

% Utilizing Automatic Speech Recognition (ASR) obtained from Whisper, OpenSmile analyzes acoustic features and detects voice activity to predict backchannel opportunities. 

% \begin{figure}[hbt!]
%     \centering
%     \includegraphics[scale=0.2]{figures/AL_Diagram.png}
%     \caption{Overview of system architecture}
%     \label{fig:sys_architecture}
% \end{figure}

% \begin{figure}
%     \centering
%     \includegraphics[width=0.95\linewidth]{figures/backchannel.png}
%     \caption{Back-channeling}
%     \label{fig:enter-label}
% \end{figure}

% \begin{figure}
%     \centering
%     \includegraphics[width=0.95\linewidth]{figures/active_listening_detection.png}
%     \caption{Active Listening Module}
%     \label{fig:enter-label}
% \end{figure}

% \subsection{Robot System}
% Furhat is a social robot platform developed to facilitate conversation between humans and technology~\cite{al2012}. Furhat can dynamically change various facial expressions through real-time projection. Furhat could demonstrate human-like behaviors from verbal to non-verbal expression (e.g., emotional expression or gaze direction). This made Furhat one of the most versatile robots in diverse conversation contexts, from multi-party to personalization~\cite{robb2023seeing, lumer2023indirect, haefflinger2023data, parreira2022design, janssens2022cool}. Thus, these advantages made Furhat a suitable robot for small and casual talks to increase the intimacy between humans and social robots.

\subsection{Sentiment-based Backchanneling Generation}
\label{emo_bc}
\subsubsection{Identifying Opportunities for Robot Backchanneling} 
We extended a rule-based backchannel opportunity prediction (BOP) model previously proposed by Park et al.~\cite{park2017backchannel, park2017telling} that used acoustic features to identify backchannel opportunities for a robot in a storytelling task with young children. Similarly to their approach, we used OpenSMILE~\cite{eyben2010opensmile} to extract audio-related characteristics, such as fundamental frequency (F0), energy, and pitch from the human interlocutor's speech in real-time. Since these models were designed based on children's prosodic cues, we fine-tuned some of the parameters to better suit adult-robot conversations based on qualitative feedback from a pilot study.% through a small pilot study with five participants.

%Specifically, we Qualitative feedback from the pilot study suggested that the robot's frequent backchanneling behavior was perceived as disruptive and inappropriate, which aligned with the findings of Li et al.~\cite{li2010backchannel}. Thus, 
Specifically, the minimal interval between each backchannel was increased to 3.0s, and the threshold for P\&P\_PAUSE (the long pause after the fluctuation of pitch frequency) was increased to 800ms,  preceded by at least 1.5s of speech. These changes reduced the frequency of our robot's backchannel compared to the original system proposed by Park et al.~\cite{park2017backchannel}.

\subsubsection{Generating Sentiment-based Backchannels} Incorporating sentiment-based backchannels that aligned with the emotional tone of the conversation could help the speaker feel more at ease, understood, and appreciated, potentially fostering a more empathetic dialogue overall~\cite{shahverdi2023}. For instance, responding ``oh no...'' to a friend who says ``I failed my algebra exam today and I feel horrible now.'' would be more appropriate than responding with a neutral backchannel, such as ``mmhmm'', which might be perceived as dismissive or unempathetic. 

In order to classify the appropriate type of robot backchannels that align with the conversational context, the Sentiment Analyzer module used the GPT-4o model to classify the sentiment of participants' utterances. We adapted the prompt developed by Kheiri et al.~\cite{kheiri2023sentimentgpt}, and classified each utterance as \textit{negative}, \textit{neutral}, or \textit{positive}. These sentiment outputs were then mapped to the corresponding backchanneling behaviors (both verbal utterances and accompanying facial expressions and gestures) of the robot.%We restrictively deprecated the explanation to constrain the model's output. The predicted sentiment would be utilized for 

\raggedbottom
\subsection{Active Listening}
\label{sec:active_listening}
\label{verbal_active_listening}
We prompted the GPT-4o model to generate empathetic responses to participants' utterances based on the principles of active listening. Previously, Welivita and Pu~\cite{welivita2024chatgpt} studied empathy-defined ChatGPT that explicitly detailed cognitive, affective, and compassionate aspects of empathy and found it significantly improved perception of empathy. Thus, we adapted and prompted an empathy-defining prompt based on Welivita and Pu~\cite{welivita2024chatgpt}: % highlighta sense of validation by not producing any judgment. Instead, we instructed the GPT-4o model to paraphrase the content and emotions of the speakers in a short manner with unconditional acceptance and constrained the output to have an average of 28 words and a maximum of 97 words.  No further questions will be asked to ensure the flow of conversation. Below is the prompt used to generate 
%  and prompted the GPT-4o model as below

\begin{verbatim}
Active Listening is a complex skill that 
involves multiple components:
• Refraining from judgment and 
paraphrasing the speaker’s message.
• Reflecting back feelings and contents.
• Demonstrating a sense of validation.
• Unconditional acceptance and unbiased 
reflection of a client’s experience
You are engaging in a conversation with a 
human. Respond in an active listening 
manner to the following using on average 
28 words and a maximum of 97 words. Do 
not ask any question.
\end{verbatim}

\begin{comment}
    \begin{tcolorbox}[colback=purple,tcbox raise base,colback=blue!3]
    Active Listening is a complex skill that involves multiple components:
    \begin{itemize}
        \item Refraining from judgment and paraphrasing the speaker's message.
        \item Reflecting back feelings and contents.
        \item Demonstrating a sense of validation.
        \item Unconditional acceptance and unbiased reflection of a client's experience
    \end{itemize}
    You are engaging in a conversation with a human. Respond in an active listening manner to the following using on average 28 words and a maximum of 97 words. Do not ask any questions
    \end{tcolorbox}
    %\caption{The GPT-4o prompt for generating the robot's active listening response.}
    \label{fig:GPT4-AL-prompt}
\end{comment}

\section{Experimental Study}

% (Section~\ref{verbal_active_listening})
% (Section~\ref{sentiment_analysis}

\subsection{Participants}
We recruited $N$=$65$ adults who were (1) 18 years old or above, (2) native or fluent in English, and (3) could travel to our local campus for the study. The recruited participants consisted of 23 females, 38 males, 2 undisclosed genders, and 2 others. Forty-six participants were between the ages of 18-24 years, seventeen participants were between 25-40 years, one was between 41-54 years, and one was above 55+ years. Out of the sixty-five participants, we removed data from two participants due to system failures and three participants due to their non-compliance with the study protocol. %Some of the behaviors considered as non-compliance were (1) not answering to the robot's questions  response to questions or stopping to answer during the entire experiment, and answered questionnaires consistently identical responses in the survey. 
The remaining sixty participants were equally assigned to the three experimental conditions: 20 for the \textit{Control} condition, 20 for the \textit{BC} condition, and 20 for the \textit{BC+AL} condition. Our study protocol was reviewed and approved by Purdue's Institutional Review Board (IRB Protocol: \#IRB-2024-707).

\begin{figure*}[t]
     \centering
     \subfigure[\textit{Control}]{\includegraphics[scale=0.109]{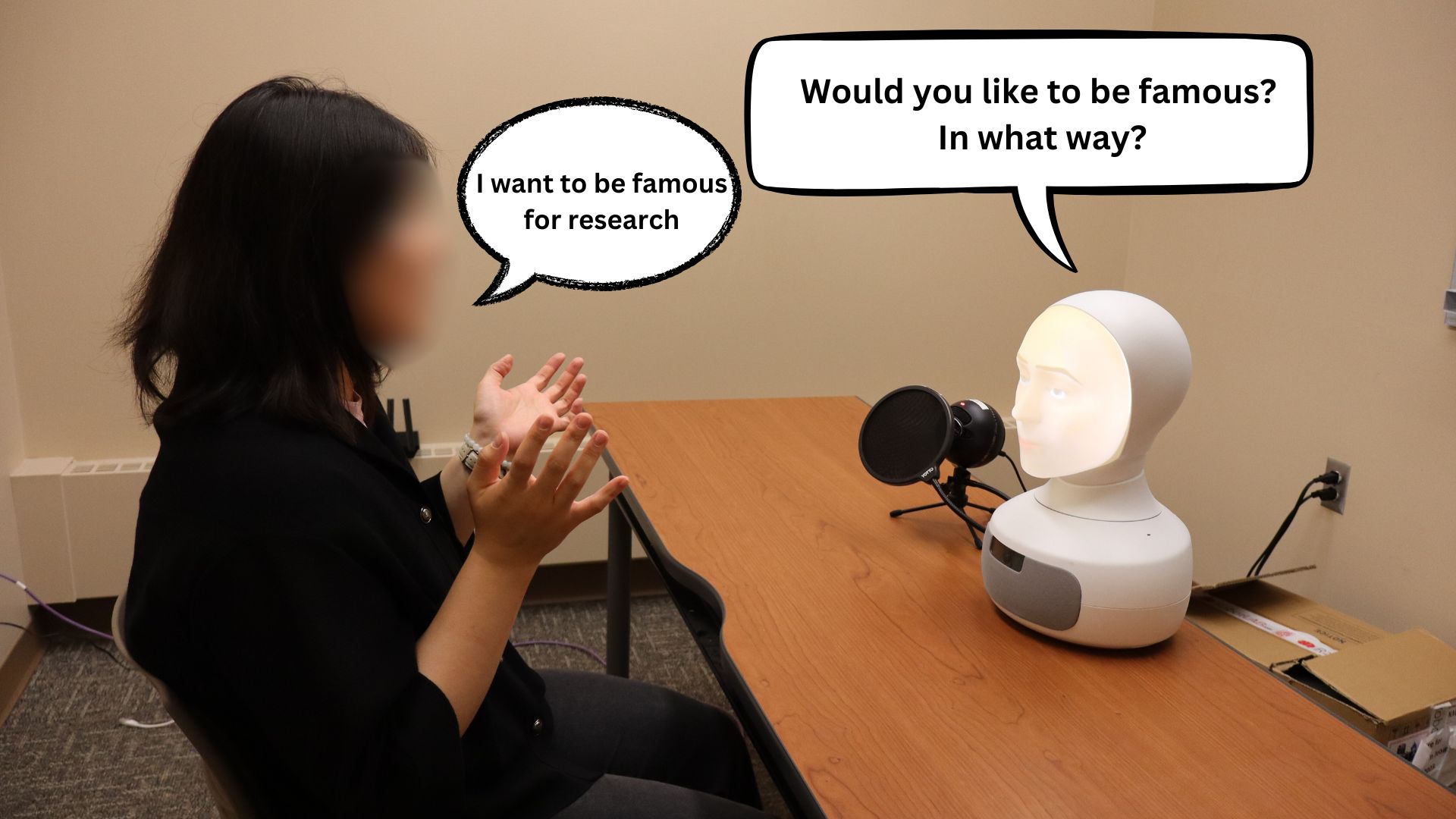}}
     \subfigure[\textit{Backchanneling-only} ]{\includegraphics[scale=0.109]{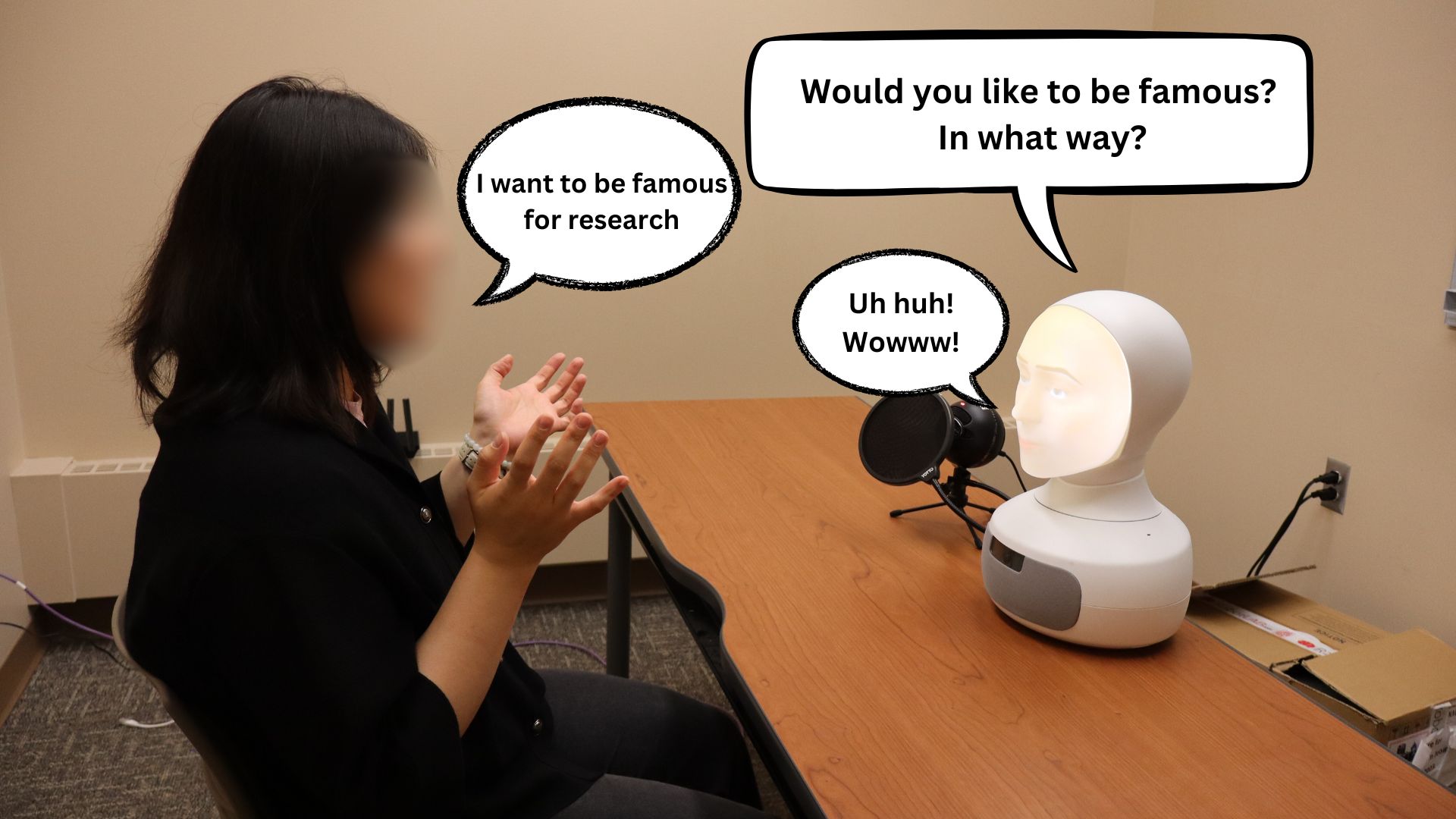}} 
     \subfigure[\textit{Backchanneling+Active Listening} ]{\includegraphics[scale=0.109]{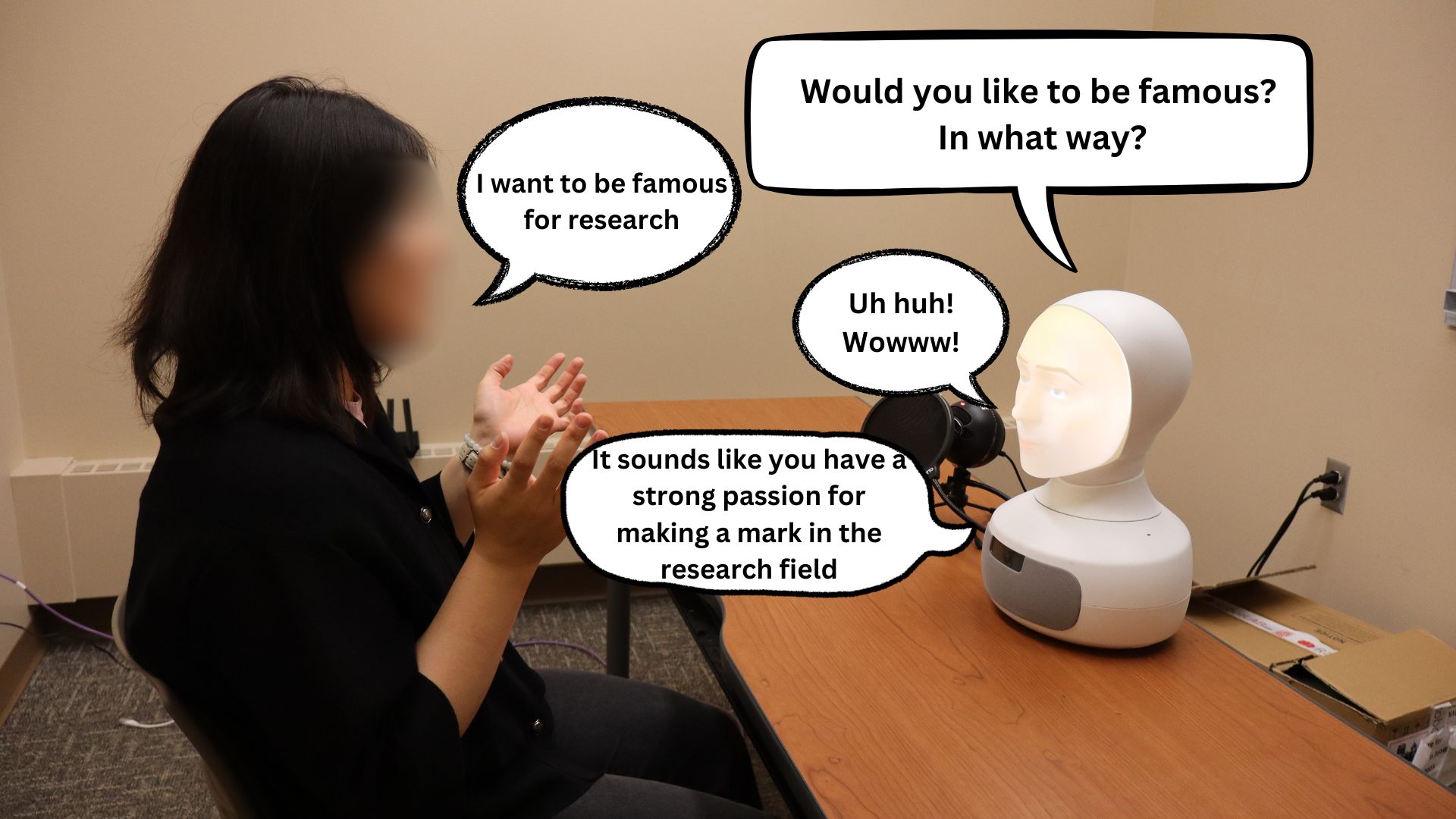}}
     \caption{We compared three robot listening behaviors. In the  (a) the control condition (\textit{Control}), the robot only asked questions, (b) the backchanneling-only condition (\textit{BC}), in which the robot produced sentiment-based backchannels while listening to the participant's response, and (c) the backchanneling+active listening behavior (\textit{BC+AL}), the robot produced sentiment-based backchannels and followed up with active listening responses.}
     \label{fig:three_condition_demo}
\end{figure*}
\subsection{Method}
\label{scripted_conversation}
%from the general public at a [anonymous] campus using physical flyers and online advertisement. 
We used physical and online advertisements to recruit study participants. Once consented, the study participants were randomized into one of the three experimental conditions (\textit{Control}, \textit{BC}, and \textit{BC+AC}), were asked to complete the demographic questionnaire, the Mini-IPIP \cite{donnellan2006}, and the Empathy Quotient \cite{baron2004} to measure their self-reported personality and empathy traits. %These measures are detailed in Section~\ref{sec:self-report-survey}.

During the study, the participant engaged in a one-on-one conversation with the robot by answering nine questions designed to elicit self-disclosures. These nine questions are a subset of thirty-six questions used in the study by Aron et al.~\cite{aron1997} and are designed to elicit self-disclosures with a gradual level of escalation and were shown to be effective in creating interpersonal closeness. %The original study included thirty-six questions with three levels of sensitivity and we chose three questions from each set to make Their study included three sets of questions that differed in the level of sensitivity The original study included thirty-six questions used as prompts for conversations that was found to result in significantly higher post-interaction closeness in comparison to other pairs who engaged in small talk~\cite{aron1997}. In their study, a total of thirty-six questions were used as conversational prompts that were divided into three sets with different intensity levels. For our study, we selected three questions from each set that demonstrated a structured experimental design that encourages strangers to foster intimacy with the robot through self-disclosure. 
%We chose nine questions  levels of sensitivity were chosen from the original thirty-six questions \cite{aron1997} for our experimental study \ref{tab:nine_questions} to elicit progressively deeper self-disclosures from the participants. % of the experimental interpersonal closeness development. We aim to make people feel understood and empathetically connected through a series of small-talk questions and self-disclosure. 
%These nine questions explored personality and evoke emotions of the participants. The conversation began with the introduction of the robot about itself. Then, the robot would ask the participant for their names (i.e. ``What is your name?'') and inquire about their well-being (i.e. ``How are you doing today?''). It also walked the participant through its interaction features, where blue LED light indicated listening and green LED light represented waiting for the response from the users. 
The first three questions (Q1-Q3) were designed for casual ice-breaking. The next three questions (Q4-Q6) explored negative (i.e. worst memory) and positive (i.e. greatest achievement) aspects of oneself. The final three questions (Q7-Q9) prompted the participants to disclose deep personal information about themselves and their relationships with others (i.e. most important item and person). Once the participants finished responding to these nine questions, they completed the post-study questionnaires and engaged in a semi-structured interview to provide qualitative feedback on their interactions with the robot.

\begin{table}[t]
\label{tab:nine_questions}
\centering
    % \resizebox{\linewidth}{!}{
    \caption{Nine questions for the self-disclosure task}
    \begin{tabular}{|c|p{7.5cm}|}
        \hline
         & \textbf{Questions} \\
         \hline
         % \multirow{3}{*}{\textbf{Casual}} & 
         {Q1} & Would you like to be famous? In what way? \\
         % & 
         {Q2} & What would constitute a ``perfect'' day for you? \\
         % & 
         {Q3} & Given the choice of anyone in the world, whom would you want as a dinner guest?  \\
         \hline
         % \multirow{3}{*}{\textbf{Moderate}} & 
         {Q4} & If a crystal ball could tell you the truth about yourself, your life, the future, or anything else, what would you want to know? \\
         % & 
         {Q5} & What is your most terrible memory? \\
         % & 
         {Q6} & What is the greatest accomplishment of your life? \\
         \hline
        % & 
        {Q7} & If you were going to become a close friend with your partner, please share what would be important for them to know. \\
         % \multirow{1}{*}{\textbf{Sensitive}} & 
         {Q8} & Imagine your house, containing everything you own, catches fire. After saving your loved ones and pets, you have time to safely make a final dash to save any one item. What would it be? Why? \\
         % & 
         {Q9} & Of all the people in your family, whose death would you find most disturbing? Why?\\
         \hline
    \end{tabular}
 %Each set has three questions. The first set (\textbf{Casual}) contains \textbf{Q1}, \textbf{Q2}, and \textbf{Q3}. The second set (\textbf{Moderate}) comprises \textbf{Q4}, \textbf{Q5}, and \textbf{Q6}. The third set (\textbf{Sensitive}) consists of \textbf{Q7}, \textbf{Q8}, and \textbf{Q9}.}
    \label{question_set}
\end{table}
\raggedbottom

\subsection{Experimental Conditions}
We designed three different listening behaviors and compared their effects on participants' perceived empathy, self-disclosure behaviors, and engagement with the robot: (1) the \textit{Control} condition, the backchanneling-only condition (\textit{BC}), and the backchanneling with active listening (\textit{BC+AL}) condition. In the \textit{Control} condition, the robot followed a pre-scripted interaction flow, in which it asked the nine questions without exhibiting any backchanneling or active listening behaviors. In the \textit{BC} condition, the robot asked the same set of questions and provided real-time sentiment-based backchannels (e.g., ``oh wow!'', ``uh-huh'', or ``oh no...'') while the participants were responding. In the \textit{BC+AL} condition, the robot asked the questions, backchanneled during the participant's speech in real-time, and responded empathetically based on the principles of active listening as discussed in Section~\ref{sec:active_listening}. If any answers failed GPT-4o's content filter, the active listening components were reverted to scripted responses to ensure the conversation's naturalness. Figure~\ref{fig:three_condition_demo} illustrates the three experimental conditions of our study. %the participant said  verbal active listening components. Utilizing the GPT-4o prompting system discussed above, the follow-up responses in the BC+AL condition captured, rephrased, and provided encouragement based on users' answers to nine questions.
Our hypotheses are as below:
\begin{itemize}
    \item \textbf{H1}: The robot's perceived empathy will be the highest in the active listening condition, followed by the backchanneling condition and the control condition (\textit{Control}$<$\textit{BC}$<$\textit{BC+AL}).
    % \item \textbf{H2}: Participants' self-disclosure will be most in amount and deepest in sensitivity in the active listening condition, followed by the backchanneling condition and the control condition. 
    \item \textbf{H2}: Participants will self-disclose the most in the active listening condition, followed by the backchanneling condition and the control condition (\textit{Control}$<$\textit{BC}$<$\textit{BC+AL}). 
    \item \textbf{H3}: Participants will report the highest level of rapport and engagement in the active listening condition, followed by the backchanneling condition and the control condition (\textit{Control}$<$\textit{BC}$<$\textit{BC+AL}).
\end{itemize}

\section{Data Collection}
\subsection{Self-reported Questionnaires}
\label{sec:self-report-survey}
Prior to the study, participants completed (1) a demographic survey, (2) the Mini-IPIP~\cite{donnellan2006}, and (3) the Empathy Quotient (EQ)~\cite{baron2004} to measure their personality traits and individual's capacity for empathy..

After the interaction with the robot, participants completed a set of post-study questionnaires: (1) conversational experience~\cite{see2019makes}, (2) the Perceived Empathy of Technology Scale (PETS)~\cite{schmidmaier2024}, (3) the Inclusion of Oneself in Another (IOS)~\cite{aron1992inclusion}, (4) the Subject Closeness Inventory (SCI)~\cite{gachter2015measuring}, and (5) the Unified Theory of Acceptance and Use of Technology (UTAUT)~\cite{heerink2010assessing}. 

The conversation experience questionnaire was on a 4-point Likert scale adapted from~\cite{see2019makes} and evaluates participants' perception of their conversations with the robot. The scale had six sub-components to evaluate the human-robot conversation: \textit{Engagingness} ``How much did you enjoy talking to this user?''; \textit{Interestingness} ``How interesting or boring did you find this conversation?''; \textit{Inquisitiveness} ``How much did the [robot] try to get to know you?''; \textit{Listening} ``How much did the [robot] seem to pay attention to what you said?''; \textit{Repetition} ``How repetitive was this [robot]?''; \textit{Sense} ``How often did this user say something which did NOT make sense?''. 

The Perceived Empathy of Technology Scale (PETS)~\cite{schmidmaier2024} was administered to evaluate how empathetic the robot was perceived. The Inclusion of Other in the Self (IOS) Scale~\cite{aron1992inclusion} and the Subject Closeness Inventory (SCI)~\cite{gachter2015measuring} were administered to measure participants' perceived closeness with the robot. Lastly, the Unified Theory of Acceptance and Use of Technology (UTAUT)~\cite{heerink2010assessing} measured participants' technology acceptance with ten sub-components in 5-point Likert scale: anxiety (ANX), attitude towards technology (ATT), facilitating conditions (FC), intention to use (ITU), perceived adaptiveness (PAD), perceived enjoyment (PENJ), perceived sociability (PS), perceived usefulness (PU), social presence (SP),  and trust (TRUST).

%The camera was placed in front of the participants to fully capture the facial expressions of the participants. The session was recorded when the participants began to fill out the pre-questionnaire and finished when the participants left the study room. The data is then uploaded to the REED folder on the University cluster. 

\subsection{Post-study Interview}
We conducted a semi-structured interview with each participant at the end of the study to gain more qualitative feedback on the interaction. The interview allowed the participants to share their experiences with the robot. Participants reported their overall experience with the study and what they liked/disliked about the interaction. We also asked for their opinions on the sets of questions, the amount of interaction, and any additional features they would like to see from the robots in the future. %The whole list of the interview questions can be founded in the Appendix, to support the replicability of this work. 

\subsection{Video and Audio Data}
Participants' interactions with the robot and their semi-structured interviews were video/audio recorded with consent. The audio data were transcribed with Microsoft Speech-to-Text and \textit{Azure Speech SDK ASR} services, which were approved institutional vendors and the research team manually checked with transcription results with the original audio data for any errors.
% and were manually checked by the research team for any errors. 
From the finalized transcriptions of each interaction data, we extracted the number of utterances generated by each participant during the study session. 

\section{Data Analysis}
%We used analyses of variance (ANOVA) and generalized linear regression models [26]. For linear regression models, the predictor variable was contrast-coded [10] as ordered values [-1 0 1], Robot, Avatar and Plush respectively for sentiment scores of verbal utterance and relational touch duration. For children's gaze behavior, the predictor variable was contrastcoded as [-1 0 1] for Plush, Robot and Avatar, respectively. For comparing the effect of agents on children's joint attention behavior, we used a chi-square test for independence.

\subsection{Statistical Analyses}
Kruskal-Wallis tests with post-hoc Dunn's tests were conducted to compare participants' responses to post-study surveys (e.g., conversational experience, PETS, IOS, SCT, and UTAUT) and behavioral outcomes across the three experimental conditions. We applied the Benjamini-Hochberg corrections to control for the false discovery rate. To test the order effect (\textit{Control}$<$\textit{BC}$<$\textit{BC+AL}) among the experimental conditions, we used contrast-coded generalized linear regression models (GLM). For GLM analyses, the predictor variable was contrast coded as ordered values $[-1\ 0\ 1]$ for \textit{Control}, \textit{BC}, and \textit{BC+AL} conditions, respectively. 

\subsection{Annotating the Self-disclosure Level}
In order to evaluate the level of self-disclosure in participants' responses, we prompted the GPT-4o model to evaluate each response based the guidelines proposed by Barak and Gluck-Ofri~\cite{barak2007degree}: \textit{Information}, how much information was demonstrated from the utterances, \textit{Thoughts}, how much thoughts were put into the answer, and \textit{Feelings}, how deep the feelings were expressed per question. All three metrics had the same scale from 1 to 3, with 1 representing the lowest level of disclosure to 3 representing the highest level of disclosure. To validate the reliability of the labels generated by the GPT-4o model, we computed inter-rater reliability among the GPT-4o model and two human annotators, measured by Fleiss' Kappa. First, two human annotators labeled a subset (54 responses, 10\% of total data) of self-disclosure responses to achieve substantial inter-rater reliability between the two annotators measured with Cohen's Kappa, $\kappa$=$0.629$, $\kappa$=$0.912$, and $\kappa$=$0.838$. Then, we evaluated the reliability of the two human annotators and the GPT-4o model via Fleiss' Kappa, and found that the GPT-4o model was able to generate reliable labels that align with human annotations for each self-disclosure annotations task: \textit{Information} $\kappa$=$0.684$, \textit{Thoughts} $\kappa$=$0.747$, and \textit{Feelings} $\kappa$=$0.661$. 

We used Affdex 2.0~\cite{bishay2023affdex} to extract participants' engagement exhibited through their facial expressions during their interactions with the robot. The \textit{Engagement} metric, also known as \textit{Expressivenss}, computes the emotional engagement of the subject in a scale from 0 to 100 by using a weighted sum of upper and lower facial muscle activation units, including inner and outer brow raises, brow furrow, cheek raise, nose wrinkle, lip corner depressor, chin raise, lip press, mouth open, lip suck, and smile.%\footnote{https://blog.affectiva.com/emotion-ai-101-all-about-emotion-detection-and-affectivas-emotion-metrics}.

% Engagement, also known as Expressiveness, measures facial muscle activation to reflect a subject’s emotional involvement. It is quantified on a scale from 0 to 100. This measure is calculated as a weighted sum of various facial expressions, including inner and outer brow raises, brow furrow, cheek raise, nose wrinkle, lip corner depressor, chin raise, lip press, mouth open, lip suck, and smile.

% The \textit{Engagement} score measures ``the level of expressiveness and involvement, reflects how actively an individual responds to stimuli''.\footnote{https://imotions.com/products/imotions-lab/modules/fea-facial-expression-analysis/} by measuring the weighted sum of upper ...[add information about facial action units ]

% \begin{itemize}
%     \item Mini-IPIP \cite{donnellan2006}: We collected responses with a 20-item short form to collect personality traits in five categories; extroversion, agreeableness, conscientiousness, neuroticism, and openness. Each question consists of five-likert scale, 
%     \item Empathy Quotient \cite{baron2004}:
%     \item Conversation evaluation in 
%     \cite{see2019makes}: 
%     \item Inclusion of oneself in another (IOS; \cite{aron1992inclusion})
%     \item Subjective Closeness Index \cite{berscheid1989relationship}
%     \item Unified Theory of Acceptance and Use of Technology (UTAUT; \cite{heerink2010assessing}) 
%     \item Perceived Empathy of Technology Scale (PETS; \cite{schmidmaier2024})
% \end{itemize}

\section{Results}
\subsection{Conversational Experience}
\begin{figure}[t]
    \centering
    \includegraphics[width=\linewidth]{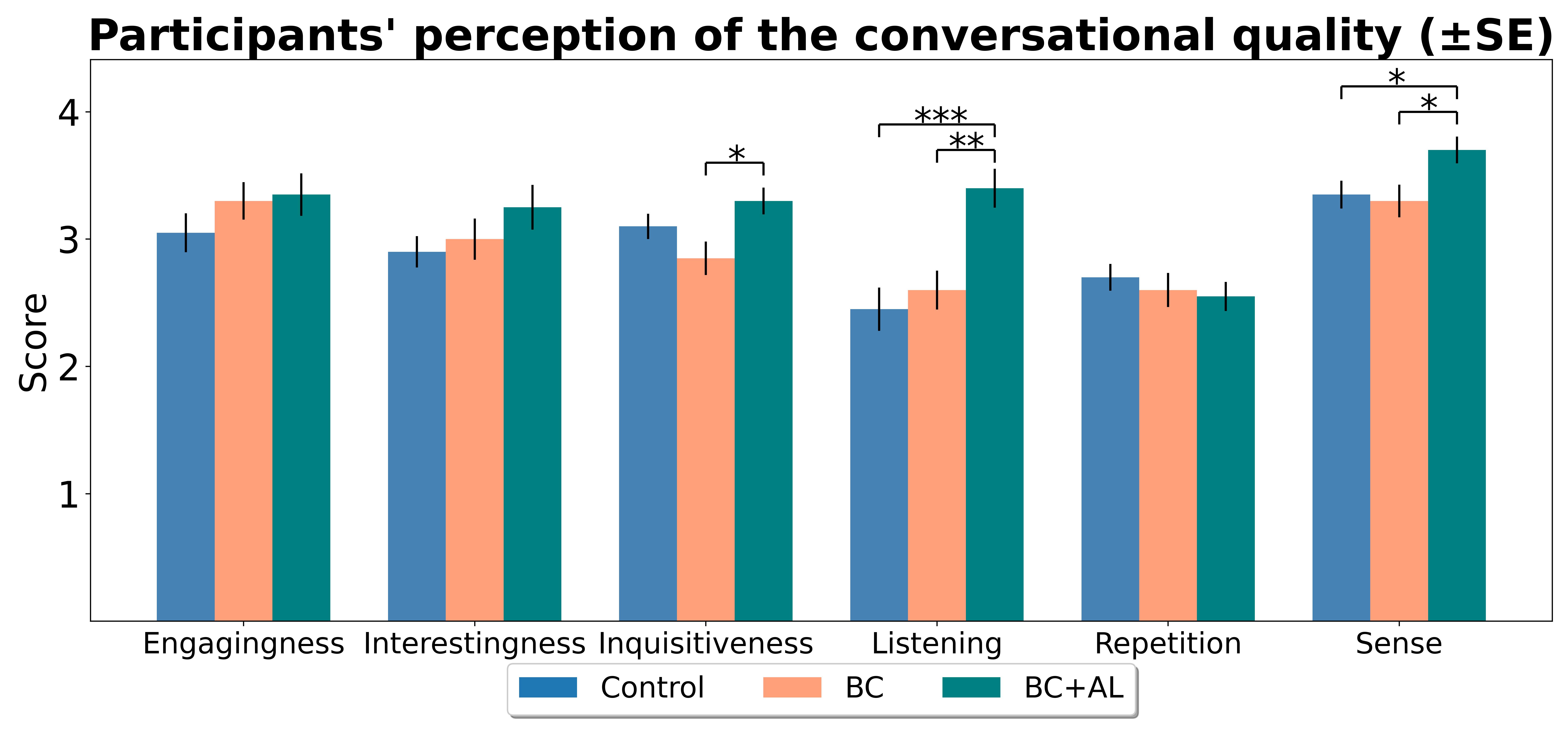}
    \caption{Participants reported statistically significant differences in their perception of the robot's \textit{Inquisitiveness}, \textit{Listening}, and \textit{Sense}) across the three experimental conditions.} 
    \label{fig:enter-label}
\end{figure}

%Table \ref{sr_question} presents detailed information on the statistical analyses for the self-reported questionnaire responses.
Among the six sub-components of the Conversational Experience questionnaire, Kruskal-Wallis tests found statistically significant differences among three of the sub-components of the Conversational Experience measures. Specifically, we found statistically significant differences in \textit{Inquisitiveness} ($\chi^2(2)$=$6.985$, $p$=$0.030$, $\varepsilon^2$=$0.118$), \textit{Listening} ($\chi^2(2)$=$15.889$, $p$=$0.0003$, $\varepsilon^2$=$0.269$), and \textit{Sense} ($\chi^2(2)$=$6.638$, $p$=$0.036$, $\varepsilon^2$=$0.112$) as represented in Figure \ref{fig:enter-label}. Post-hoc Dunn's tests further revealed that for \textit{Inquisitiveness}, only the \textit{BC+AL} condition exhibited a statistically significant difference compared to \textit{BC} ($p$=$0.024$). No significant differences were observed between \textit{BC+AL} and \textit{Control} ($p$=$0.226$) or between \textit{BC} and \textit{Control} ($p$=$0.226$). For \textit{Listening}, the post-hoc Dunn's tests indicated statistically significant differences between \textit{BC+AL} and \textit{BC} ($p$=$0.003$) as well as between \textit{BC+AL} and \textit{Control} ($p$=$0.0005$). However, no significant difference was found between \textit{BC} and \textit{Control} ($p$=$0.493$). Similarly, for \textit{Sense}, significant differences were observed between \textit{BC+AL} and \textit{BC} ($p$=$0.048$) and between \textit{BC+AL} and \textit{Control} ($p$=$0.048$), whereas no significant difference was found between \textit{BC} and \textit{Control} ($p$=$0.868$).

Furthermore, contrast-coded Generalized Linear Models (GLM) revealed a trend (\textit{Control}$<$\textit{BC}$<$\textit{BC+AL}) in both \textit{Listening} and \textit{Sense}. The analysis showed an increasing trend in \textit{Listening} ($F(1, 58)$=$4.180$, $p$=$0.000$, $f^2$=$0.338$) and a similar increasing trend in \textit{Sense} ($F(1, 58)$=$2.132$, $p$=$0.033$, $f^2$=$0.078$). However, no significant trend was observed for \textit{Inquisitiveness} ($F(1, 58)$=$1.197$, $p$=$0.231$, $f^2$=$0.024$).

\subsection{Other Self-reported Measures}
As shown in Table~\ref{sr_question}, we did not find any statistically significant differences in participants' perceived robot empathy (PETS), closeness to the robot (IOS and SCI), or technology acceptance (UTAUT) based on the Kruskal-Wallis test and contrast-coded GLM analyses.

\begin{table}[t]
\centering
\caption{Participants' self-reported conversational experience, perceived robot empathy (PETS), closeness (IOS, SCI), and technology acceptance (UTAUT).}
\scalebox{0.63}{
    \begin{tabular}{|c|c|c|c|c|c|c|}
        \hline
         & &  \multicolumn{3}{|c|}{Median} & \multicolumn{2}{|c|}{Kruskal-Wallis Test} \\
         \hline
         Questionnaire & Sub-component & \textit{Control} & \textit{BC} & \textit{BC+AL} & Chi-squared & p-value \\
         \hline
         & Engagingness & 3.05($\pm 0.686$) & 3.3($\pm 0.656$) & 3.35($\pm 0.745$) & 2.296 & 0.317\\
         & Interestingness & 2.9($\pm 0.552$) & 3.0($\pm 0.725$) & 3.25($\pm 0.786$) & 2.811 & 0.245  \\
         Conversation & Inquisitiveness & 3.1($\pm 0.447$) & 2.85($\pm 0.587$) & 3.3($\pm 0.470$) & \textbf{6.985} & \textbf{0.030*}  \\
         Experience& Listening & 2.45($\pm 0.759$) & 2.6($\pm 0.680$) & 3.4($\pm 0.680$) & \textbf{15.889} & \textbf{0.000***}  \\
         $M$($SD$) & Repetition & 2.7($\pm 0.470$) & 2.6($\pm 0.598$) & 2.55($\pm 0.510$) & 0.225 & 0.893  \\
         & Sense & 3.35($\pm 0.489$) & 3.3($\pm 0.571$) & 3.7($\pm 0.470$) & \textbf{6.638} & \textbf{0.036*} \\
         \hline
         & PETS\_ER & 51.166 & 49.916 & 60.916 & 2.248 & 0.324 \\
         PETS & PERS\_UT & 40.25 & 47.125 & 51.0 & 2.249 & 0.324\\
         & PETS & 45.6 & 45.8 & 52.75 & 2.292 & 0.317 \\
         \hline
         IOS & IOS & 2.0 & 2.0 & 2.0 & 0.093 & 0.954  \\ 
         \hline
         SCI & SCI\_1 & 2.0 & 2.0 & 2.0 & 0.822 & 0.662  \\
             & SCI\_2 & 2.0 & 2.0 & 2.0 & 0.616 & 0.734  \\
         \hline
         & ANX & 2.0 & 2.5 & 2.625 & 1.636 & 0.441  \\
         & ATT & 3.833 & 3.333 & 3.666 & 1.260 & 0.532    \\
         & FC & 3.0 & 3.25 & 3.5 & 0.849 & 0.654   \\
         & ITU & 2.0 & 2.0 & 2.0 & 2.505 & 0.285   \\
         UTAUT & PAD & 3.333 & 3.666 & 3.666 & 1.769 & 0.412   \\ 
         & PENJ & 3.4 & 3.5 & 3.4 & 0.005 & 0.997  \\
         & PS & 3.375 & 3.375 & 3.0 & 0.446 & 0.799  \\
         & PU & 2.666 & 3.0 & 2.5 & 2.548 & 0.279  \\
         & SP & 2.875 & 2.75 & 2.5 & 4.133 & 0.126   \\
         & TRUST & 3.0 & 3.0 & 2.5 & 2.344 & 0.309  \\
         \hline
    \end{tabular}
}
    % \vspace{-8pt}
    \label{sr_question}
\end{table}

\begin{figure*}[ht]
    \centering
    \subfigure[Number of utterances across three conditions\label{fig:num_utterance_total}]{\includegraphics[width=0.32\textwidth]{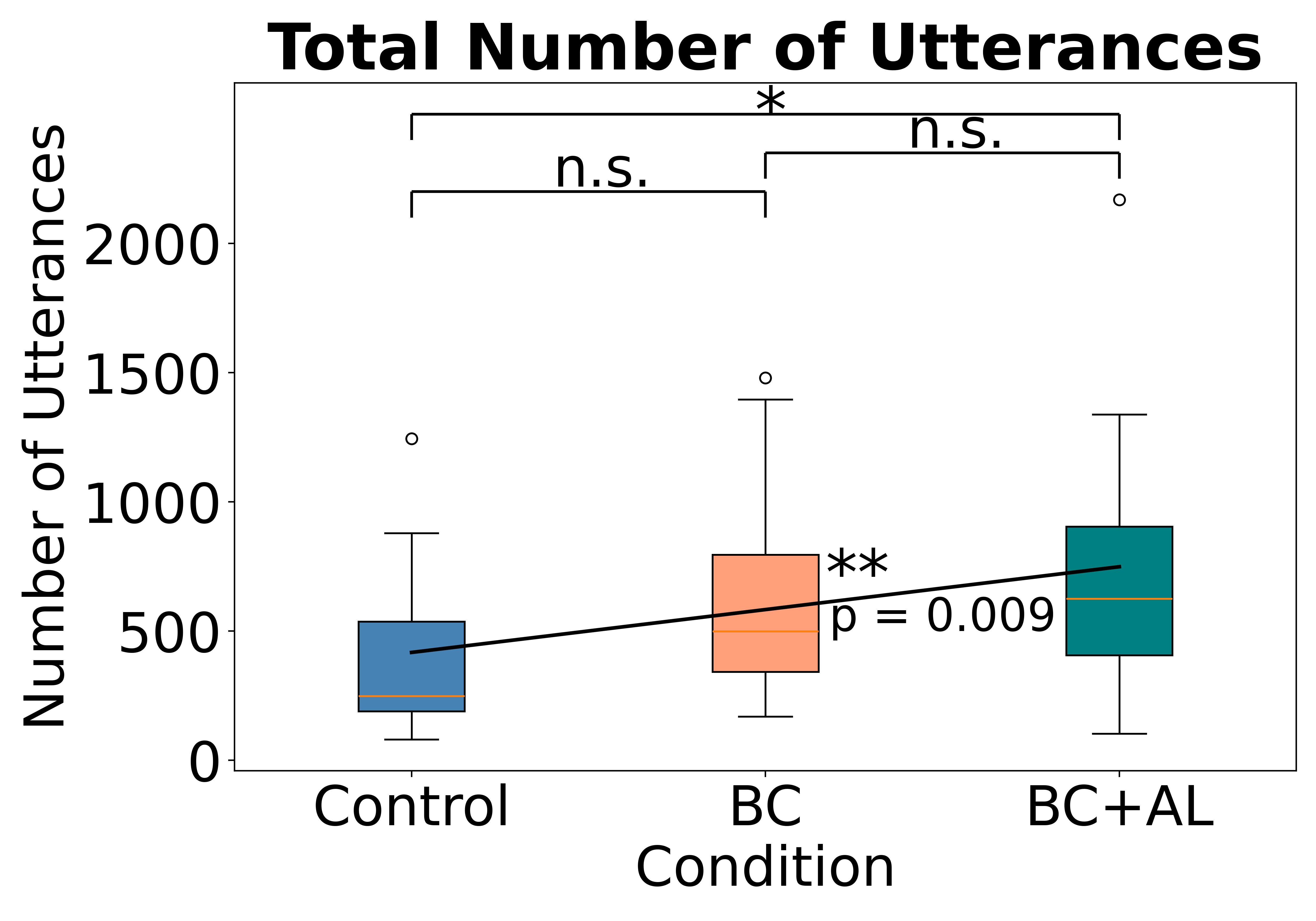}} 
    \subfigure[\textit{Self-disclosure level} across three conditions.\label{fig:ans_eval}]{\includegraphics[width=0.32\textwidth]{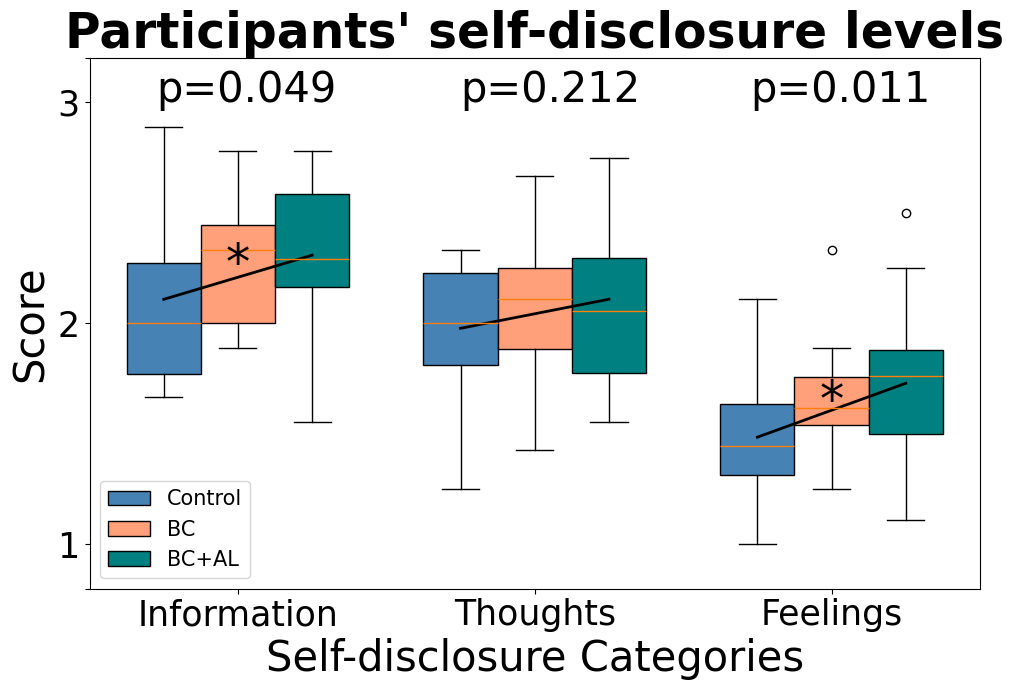}}
    %\subfigure[\textit{Sentimentality} scores across three conditions.]{\includegraphics[width=0.32\textwidth]{figures/sentimentality.jpg}}
    \subfigure[\textit{Engagement} scores across three conditions.\label{fig:facial_exp}]{\includegraphics[width=0.315\textwidth]{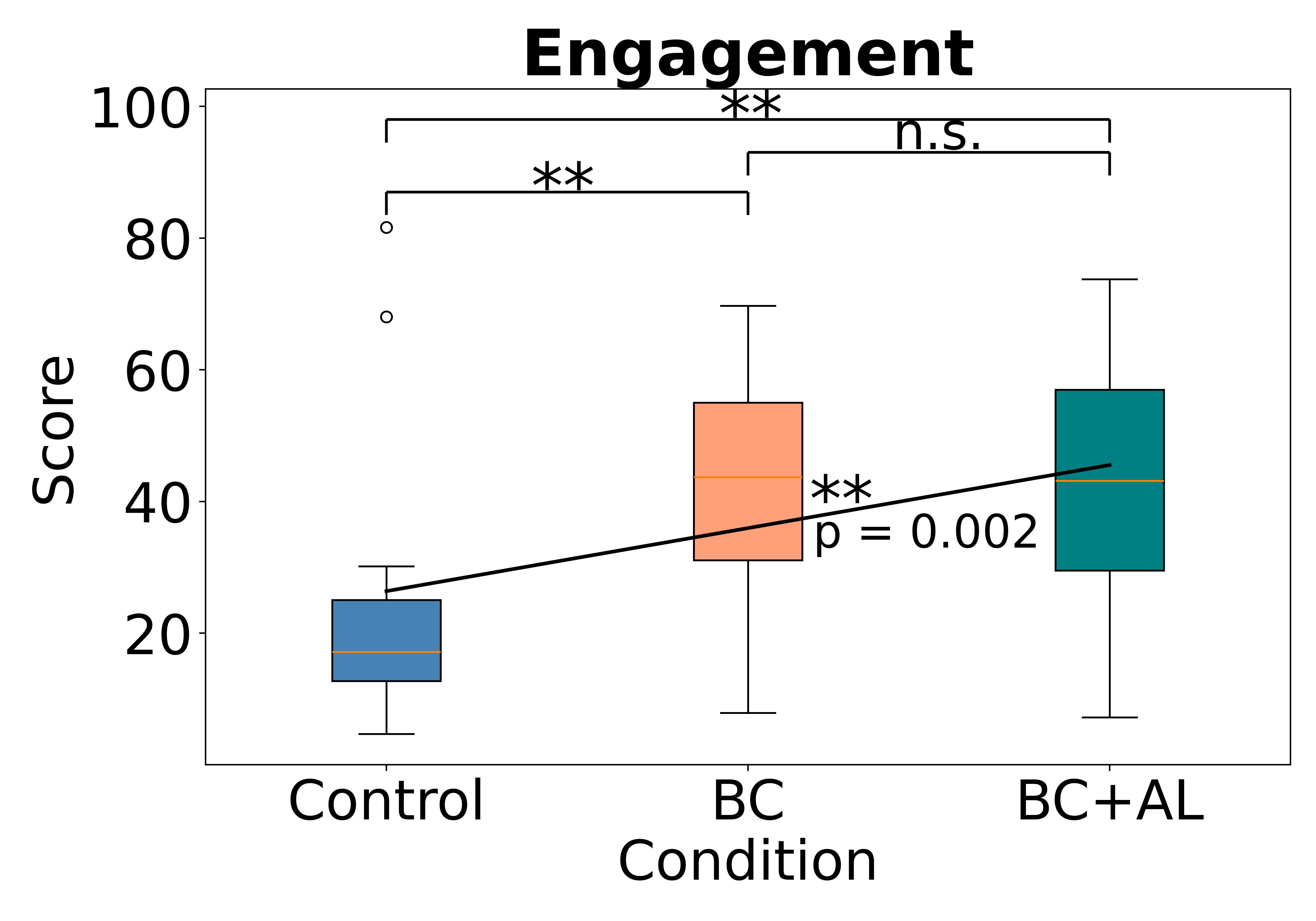}}
    
    \caption{We found that increasing trends of \textit{Control}$<$\textit{BC}$<$\textit{BC+AL} in (a) the number of participants' verbal utterances, (b) the level of two types of self-disclosures (\textit{Information} and \textit{Feelings}) in participants' responses, and (c) the level of \textit{Engagement} observed in participants' facial expressions. %increase trends across verbal analysis (i.e. the total number of utterances) and facial expression analysis (i.e. \textit{Sentimentality} and \textit{Engagement}). The \textit{BC+AL} condition showed a statistical significance compared to \textit{Control} condition in all three metrics, while \textit{BC} only demontrated the statistical significance in \textit{Engagement} measurement.
    }
    \vspace{-8pt}
\end{figure*}

\raggedbottom

\subsection{The Number of Verbal Utterances}
%The mean of a contrast-coded GLM was conducted to test the order 417.61 (SD 308.87), 537.72 (SD 300.67), and 764.05 (SD 509.64) for Control, BC, and BC+AL, respectively. 
Figure~\ref{fig:num_utterance_total} shows an increased trend in the number of utterances between conditions. Kruskal-Wallis tests revealed that the number of total verbal utterances showed statistically significant differences across the three experimental conditions: \textit{Control} $MD$=$249.0$, \textit{BC} $MD$=$499.5$, and \textit{BC+AL} $MD$=$625.5$, $\chi^2(2)$=$7.606$, $p$=$0.022$, $\varepsilon^2$=$0.128$. The number of utterances was significantly higher in the \textit{BC+AL} condition than the \textit{Control} condition ($p$=$0.025$), but there was no difference between the \textit{BC+AL} condition and the \textit{BC} condition ($p$=$0.541$), or the \textit{BC} condition and the \textit{Control} condition ($p$=$0.064$). The contrast-coded GLM showed that there was a statistically significant trend of increase in the number of utterances (\textit{Control}$<$\textit{BC}$<$\textit{BC+AL}) ($F(1, 58)$=$2.631$, $p$=$0.009$, $f^2$=$0.122$).

% \begin{figure}[ht!]
%     \centering
%     \includegraphics[width=0.8\linewidth]{figures/total_utterance.jpg}
%     \caption{The total number of utterances across conditions.}
%     \label{fig:num_utterance_total}
% \end{figure}

\subsection{The Level of Self-disclosures}
We did not find any statistically significant differences in participants' self-disclosure levels via Kruskal-Wallis tests: \textit{Information} ($\chi^2(2)$=$5.747$, $p$=$0.056$, $\varepsilon^2$=$0.097$), \textit{Thoughts} ($\chi^2(2)$=$1.379$, $p$=$0.501$, $\varepsilon^2$=$0.023$), and \textit{Feelings} ($\chi^2(2)$=$3.822$, $p$=$0.147$, $\varepsilon^2$=$0.104$). The medians for \textit{Information} were: \textit{Control} $MD$=$2.0$, \textit{BC} $MD$=$2.333$, and \textit{BC+AL} $MD$=$2.291$. The medians for \textit{Thoughts} were: \textit{Control} $MD$=$2.0$, \textit{BC} $MD$=$2.111$, and \textit{BC+AL} $MD$=$2.055$. The medians for \textit{Feelings} were: \textit{Control} $MD$=$1.444$, \textit{BC} $MD$=$1.619$, and \textit{BC+AL} $MD$=$1.763$. However, the contrasted-coded GLM analysis showed that there was statistically significant trends of increase in \textit{Information} ($F(1, 58)$=$1.971$, $p$=$0.049$, $f^2$=$0.067$) and \textit{Feelings} ($F(1, 58)$=$1.999$, $p$=$0.046$, $f^2$=$0.114$). We did not find any significant trend in \textit{Thoughts} scores ($F(1, 58)$=$1.248$, $p$=$0.212$, $f^2$=$0.026$) as represented in Figure~\ref{fig:ans_eval}.   

\begin{comment}
    
\begin{figure}[h]
    \centering
    \includegraphics[width=0.8\linewidth]{figures/three_levels.png}
    \caption{Three self-disclosure categories (i.e. \textit{Information}, \textit{Thoughts}, and \textit{Feelings}) did not show statistically significant difference between the three experimental conditions. However, \textit{Information} and \textit{Feelings} demonstrated an increase trend across three conditions.}
    \label{fig:ans_eval}
\end{figure}
\end{comment}

% Add example of 3 and 1 of Information from participants

\subsection{Facial Expressions During the Interaction}
%Kruskal-Wallis tests demonstrated that statistical significant difference across three conditions \textit{Sentimentality} ($H$=$9.373$, $p$=$0.009$) and  For the \textit{Sentimentality} score, the \textit{BC+AL} and \textit{Control} showed statistically significant difference  ($p$=$0.006$) but there was no difference between \textit{BC+AL} and \textit{BC} ($p$=$0.147$); and \textit{BC} and \textit{Control} ($p$=$0.147$). 
Results from the Kruskal-Wallis test revealed that there was a statistically significant difference among participants' \textit{Engagement} exhibited in their facial expressions as represented in Figure~\ref{fig:facial_exp}:  \textit{Control} $MD$=$17.19$, \textit{BC} $MD$=$43.71$, \textit{BC+AL} $MD$=$43.14$, $\chi^2(2)$=$14.052$, $p$=$0.0008$, $\varepsilon^2$=$0.238$. The post-hoc Dunn's tests showed that the \textit{Engagement} scores for the \textit{Control} condition were significantly lower than the \textit{BC} condition, $p$=$0.002$; and the \textit{BC+AL} condition, $p$=$0.002$; but not between \textit{BC+AL}-\textit{BC}, $p$=$0.935$. However, the contrast-coded GLM revealed that participants' facial expressions showed a statistically significant increase in \textit{Engagement} scores across the three conditions: \textit{Control}$<$\textit{BC}$<$\textit{BC+AL}, $F(1,58)$=$3.165$, $p$=$0.002$, $f^2$=$0.182$. %Thus, participants tend to express and interact more in \textit{BC+AL}, followed by \textit{BC} and \textit{Control} condition.

\subsection{Post-Study Interview}
\label{result_post_study}

In general, participants reported that conversations with the active listening robot (\textit{BC+AL} condition) were more natural and comfortable. %, followed by \textit{BC} and \textit{Control} condition. 
Participants in the \textit{BC} and \textit{Control} conditions reported that the robot's usage of language did not feel human-like. For instance, P32 in the \textit{Control} condition mentioned ``\textit{It seems to have programmed responses. And I am not sure if he [social robot] listened to what I was saying because a few times I said something and he [social robot] responded in the opposite way or heard it in the opposite way}.'' Similarly, P07 in the \textit{BC} condition said that ``\textit{It [social robot] wasn't very human. It is very robotic and didn't really seem to pay attention to what I said. And the comments it gave while I was speaking were awkward. ... It wasn't that natural}.'' However, people in the \textit{BC+AL} condition were more forgiving of the robot's mistakes (e.g., inappropriate backchanneling timing or the lack of elaboration in personalized questions). P03 in the \textit{BC+AL} condition said ``\textit{I think it's really interesting. I really enjoyed talking with him because his voice is really clear and the usage of vocabulary is very impressive. ... Although there were some pauses when I delivered my speech, I still think it is acceptable because the responses were really amazing, so the pauses were acceptable}.'' Also, P18 said ``\textit{It felt like it was actually trying to get to know me. Um, it was asking insightful questions, and um it seems like it was listening to my answers and responding. Then it also seemed to elaborate on my answers. However, it would probably be more interesting if it could ask some more personal questions}.''

\section{Discussions}
We developed a social robot that can exhibit sentiment-based backchanneling and active listening behavior to elicit human interlocutors' self-disclosures during a conversational task. Through our experimental study and participants' self-reported responses, we found that participants who interacted with the active listening robot (\textit{BC+AL}) rated their conversations with the robot to be more inquisitive (\textit{Inquisitiveness}) and sensible (\textit{Sense}) and as a better listener (\textit{Listening}) than the robot that only backchanneled (\textit{BC}) or the robot that did not show any explicit listening behaviors (\textit{Control}). 

%The results of the qualitative analysis suggest that the robot's active listening response did not always align with the questions and responses of the participants, and this could have been caused by a number of factors. Firstly, context plays a big role in forming the appropriate active listening answers. Since we did not add the questions as context for GPT-4o models, combined with short and lack of context answers causes inappropriate answers. Secondly, the Whisper Speech-to-Text engine often incorrectly transcribed participants' utterances and led to erroneous or inappropriate sentiment-based backchannels and robot responses.  %This failure not only damages the active listening response but also harm sentiment analysis, where emotional-specific backchannel behaviors did not match with the responses from the participants. Thus, this errors causes GPT to give proper response, which fail to express empathetic towards the participants. 

In addition, results from the behavioral analyses showed that participants elicited more amount and deeper quality of self-disclosures and were visually more engaged with their facial expressions when conversing with an active listening robot (\textbf{H2}).  %followed by backchanneling and active listening. 
All questions in the interaction were designed to promote self-disclosure from the speakers, and yet we found that the total number of utterances by participants was highest in the \textit{BC+AL} condition, followed by the \textit{BC} condition and then by the \textit{Control} condition. We also found a statistically significant trend of increase in the depth of \textit{Information} and \textit{Feelings} in participants' self-disclosures in the order of (\textit{Control}$<$\textit{BC}$<$\textit{BC+AL}) as defined by~\cite{barak2007degree}. Furthermore, participants in \textit{BC+AL} condition expressed a higher degree of engagement in the conversation (\textit{Engagement}) through their facial expressions, followed by \textit{BC} and \textit{Control} conditions. Thus, we found evidence to support \textbf{H3}, where active listening shows engagement with signs of building rapport.

However, we did not find any significant differences in participants' perceived empathy measured by PETS (\textbf{H1}) or closeness measured by the IOS and the SCI (\textbf{H3}) across the three experimental conditions. In general, participants reported relatively low scores (2 out of 7) regarding their perceived closeness with the robot for all IOS and SCI measures. One potential reason for this outcome could be that we only chose a subset of the thirty-six prompts/questions used in the original study \cite{aron1997}, and there was not enough time and conversations for the robot to build closeness with the participants. Although we do not know for sure that including all thirty-six questions and longer conversations would result in different self-report outcomes, we found behavioral evidence that the robot's active listening feature leads to deeper and more human self-disclosures. While there exist many ethical considerations to be made when eliciting sensitive or personal disclosures from people, a robot's ability to make people feel more willing and at ease to share about themselves could further enhance its supportive capabilities by quickly learning about each individual's unique needs, preferences, and context.

\section{Conclusion}
In this work, we presented an autonomous conversational robot that can produce sentiment-based backchanneling and active listening behavior. Our experimental study found that the active listening robot is perceived more positively by human interlocutors and is more efficacious in eliciting self-disclosures and expressive non-verbal behaviors. Qualitative feedback from the post-study interview also suggests that the robot's tailored responses and follow-up questions made participants perceive that the robot made an effort to get to know them. These results highlighted the importance of designing socio-emotionally intelligent listening behaviors for robots to enhance people's engagement and willingness to share more about themselves, which would further enhance the robot's capability to identify unique needs and preferences more efficiently. %\changes{In future work, we will refine backchannel timing and end-of-speech detection where our system sometimes fail to detect to further enhance our agent listening behavior.} %Our future work will focus on utilizing our and active listening behavior  our work offers valuable insights for social robots to provide personalized and tailored support in long-term interaction context, such as healthcare and eldercare. %We plan to further investigate ways to design better listening behavior that can elicit  between humans and robots.

\vspace{12pt}
% \color{red}

% \end{thebibliography}
% \bibliographystyle{ROMAN 2025/IEEEtran.bst}

\bibliographystyle{IEEEtran}
\bibliography{IEEEabrv, references}

\end{document}